\documentclass[a4paper]{jpconf}
\usepackage{graphicx}
\begin{document}
\title{Event-by-event fluctuations of mean transverse momentum in Pb--Pb and Xe--Xe collisions with ALICE}

\author{Tulika Tripathy (for the ALICE collaboration)}

\address{Indian Institute of Technology Bombay, Mumbai, India}

\ead{tulika.tripathy@cern.ch}

\begin{abstract}
Event-by-event fluctuations of the mean transverse momentum of charged particles produced in Pb--Pb and Xe--Xe collisions at $\sqrt{s_{\rm{NN}}}$ = 5.02 TeV and $\sqrt{s_{\rm{NN}}}$ = 5.44 TeV, respectively, are studied as a function of the charged-particle multiplicity using the ALICE detector at the LHC. Dynamical fluctuations are observed  in both collision systems which indicate correlated particle emission. The central Pb--Pb and Xe--Xe collisions show a significant reduction of the fluctuation in comparison to peripheral collisions and are in qualitative agreement with previous measurements  in Pb--Pb  collisions at $\sqrt{s_{\rm{NN}}}$ = 2.76 TeV. The results are compared with the HIJING model. A clear deviation from simple superposition of independent nucleon-nucleon collisions scenario, where the final state particles are produced from superposition of particle emitting sources, is observed.
\end{abstract}

\section{Introduction}

Fluctuations of various observables in ultrarelativistic heavy-ion collisions have been extensively studied as they provide important indications to the formation of a quark-gluon-plasma (QGP). The collisions made at the LHC produce a large number of particles, even in a single event. Such large statistics in a single collision could be used to study the fluctuations of different observables, such as mean transverse momenta, on an event-by-event basis. These fluctuation analyses can also reveal collective effects and the onset of thermalization of a system. Thus, they are proposed as one of the key observables for the investigation of the hot and dense matter generated in heavy-ion collisions~\cite{Shuryak:1997yj,Stephanov:1998dy,Stephanov:1999zu}.

Here, we report on the event-by-event fluctuations of the mean transverse momentum, $\langle p_{\rm{T} }\rangle$ in Pb-Pb collisions at $\sqrt{s_{\rm{NN}}}$  = 5.02 TeV and Xe-Xe collisions at  $\sqrt{s_{\rm{NN}}}$= 5.44 TeV.  The analysis focuses on non-statistical fluctuations of $\langle p_{\rm{T} }\rangle$ on an event-by-event basis which is represented by the two particle correalator, $\langle \Delta {p_{\rm{T,i} }},\Delta{ p_{\rm{T,j} }}\rangle$.
 
\section{Analysis details} 

The reported results in this proceedings are obtained from Pb-Pb collisions at $\sqrt{s_{\rm{NN}}}$ = 5.02 TeV and Xe-Xe collisions at $\sqrt{s_{\rm{NN}}}$ = 5.44 TeV recorded with the ALICE detector~\cite{ALICE:2008ngc} during Run 2 of the LHC. A minimum-bias trigger condition is used for both Pb--Pb and Xe--Xe collisions. The centrality is defined based on the total charge  deposited in both forward V0A and V0C detectors. The main detectors used for tracking and estimating the position of the collision vertex along the beam direction, $V_{\rm z}$, are the Inner Tracking System (ITS) and the Time Projection Chamber (TPC). Only events with at least one charged track contributing to the reconstruction of $V_{\rm z}$, and with $|V_{\rm z}|$ within 10 cm from the nominal interaction point, are accepted. To ensure uniform tracking efficiency within the TPC, we restrict the acceptance to $|\eta| < $ 0.8. The tracking efficiency drops at very low transverse momentum, therefore we only consider tracks with $p_{\rm T}$ greater than 0.15 GeV/$c$. At the same time, we restrict the analysis to the domain of soft physics by requiring track $p_{\rm T}$ lower than 2.0 GeV/$c$.

\subsection{Two-particle correlator}

The mean transverse momentum of an event, $\langle p_{\rm T} \rangle$, is calculated on an event-by-event basis as 

\begin{eqnarray}
\langle p_{\rm T} \rangle= {\frac{\sum\limits_{i=1}^{N_{\rm{ch,k}}}{p_{\rm{Ti}}}}{N_{\rm{ch,k}}}},
\end{eqnarray}
where $p_{\rm{Ti}}$ is the transverse momentum of $i^{\rm th}$ particle in the event $k$ and $N_{\rm{ch,k}}$ is the charged-particle multiplicity of the event. We further calculate the $\langle\langle p_{\rm{T}} \rangle\rangle$ by averaging the $\langle p_{\rm{T}} \rangle$ over events within the same multiplicity bin. 


The two-particle correlator is given by:
\begin{eqnarray}
\langle \Delta p_{\rm Ti}\Delta p_{\rm Tj} \rangle= \bigg\langle\frac{\sum_{i,j\neq i}(p_{\rm Ti}-\langle\langle p_{\rm T} \rangle\rangle)(p_{\rm Tj}-\langle\langle p_{\rm T} \rangle\rangle)}{ N_{\rm{ch}}(N_{\rm{ch}}-1)}\bigg\rangle.
\end{eqnarray}
Here, $p_{\rm{i}}$ and $p_{\rm{j}}$ are the transverse momenta of $i^{\rm th}$ and $j^{\rm th}$ particle of an event in a particular multiplicity bin, where $N_{\rm{ch}}(N_{\rm{ch}}-1)$ represents  the number of particle pairs.
The  above equation can be further simplified and can be rewritten as,

\begin{eqnarray}
 \langle \Delta p_{\rm Ti}\Delta p_{\rm Tj} \rangle= \bigg\langle\frac{(Q_1)^2-Q_2}{N_{\rm{ch}}(N_{\rm{ch}}-1)}\bigg\rangle - \bigg\langle\frac{Q_1}{N_{\rm{ch}}}\bigg\rangle^2,
\end{eqnarray}
where,
\begin{eqnarray}
Q_{\rm n}=\sum_{i=1}^N (p_i)^n.
\end{eqnarray}


\section{Results and discussion}

\begin{figure}[h]
\begin{center}
\includegraphics[width=16pc]{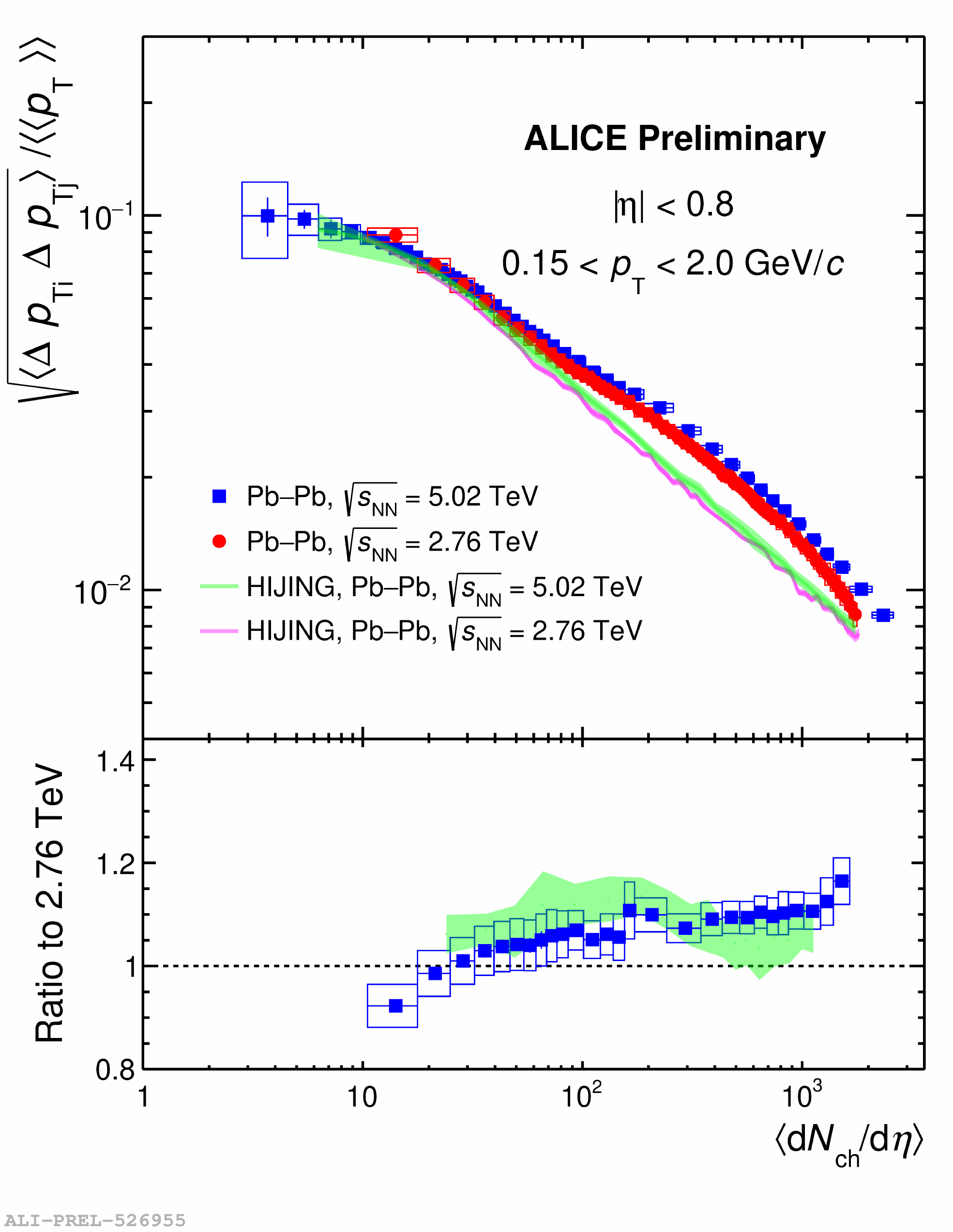}
\includegraphics[width=16pc]{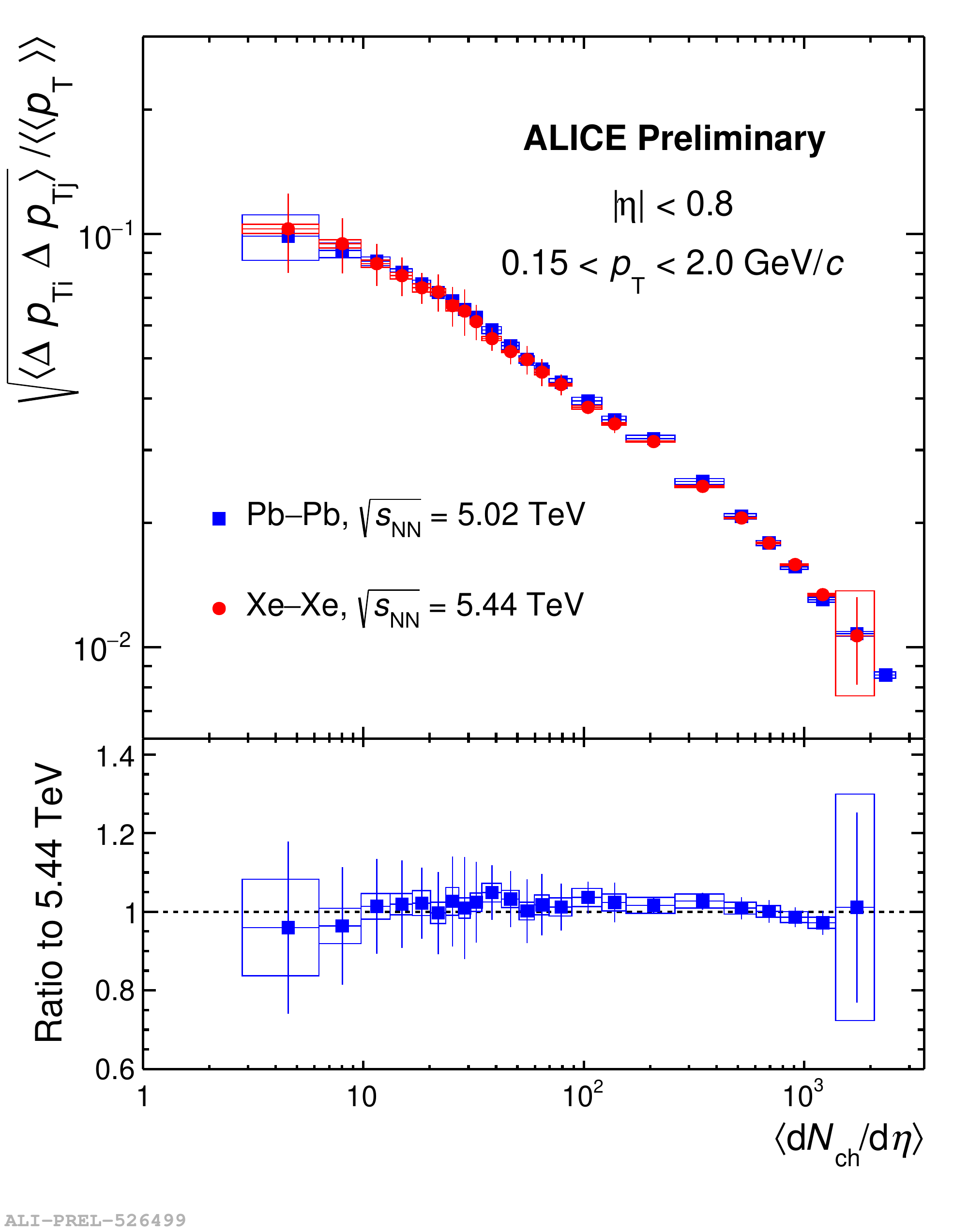}
\caption{\label{fig1} Left panel: Event-by-event fluctuations of mean transverse momentum in Pb--Pb collisions at 2.76 and 5.02 TeV and their model comparisons with HIJING, right panel: event-by-event fluctuations of mean transverse momentum in Pb--Pb collisions at $\sqrt{s_{\rm{NN}}}$ = 5.02 TeV and Xe--Xe collisions at $\sqrt{s_{\rm{NN}}}$ = 5.44 TeV}
\end{center}
\end{figure}

 The left panel of Fig.~\ref{fig1} shows the event-by-event fluctuations of $\langle p_{\rm{T} }\rangle$ in Pb--Pb collisions at $\sqrt{s_{\rm{NN}}}$ = 2.76~\cite{ALICE:2014gvd} and 5.02~TeV, represented by the dimensionless quantity $\sqrt{\langle\Delta p_{\rm{Ti}}\Delta p_{\rm{Tj}}}\rangle/\langle\langle p_{\rm{T} }\rangle\rangle$. A significant dynamical fluctuation is observed for both collision energies, which decreases with
increasing multiplicity. 

One of the key effects in Pb-Pb collisions is the stronger decrease of the $\langle p_{\rm{T} }\rangle$ fluctuations in high multiplicity bins which may be related to radial flow and other final state effects. The central Pb-Pb collisions show a significant reduction of the fluctuation in comparison to peripheral collisions and the value of the correlator is found to be higher for Pb--Pb collisions at $\sqrt{s_{\rm{NN}}}$ = 5.02 TeV compared to the previous measurements in Pb--Pb collisions at $\sqrt{s_{\rm{NN}}}$ = 2.76 TeV. The qualitative difference in the trend from the HIJING model~\cite{Deng:2010mv} and data indicates a deviation from a simple superposition scenario, where the final state particles are produced from the superposition of particle emitting sources. However, HIJING model reproduces the collision energy dependence as observed in data. 

\begin{figure}[h]
\begin{center}
\includegraphics[width=16pc]{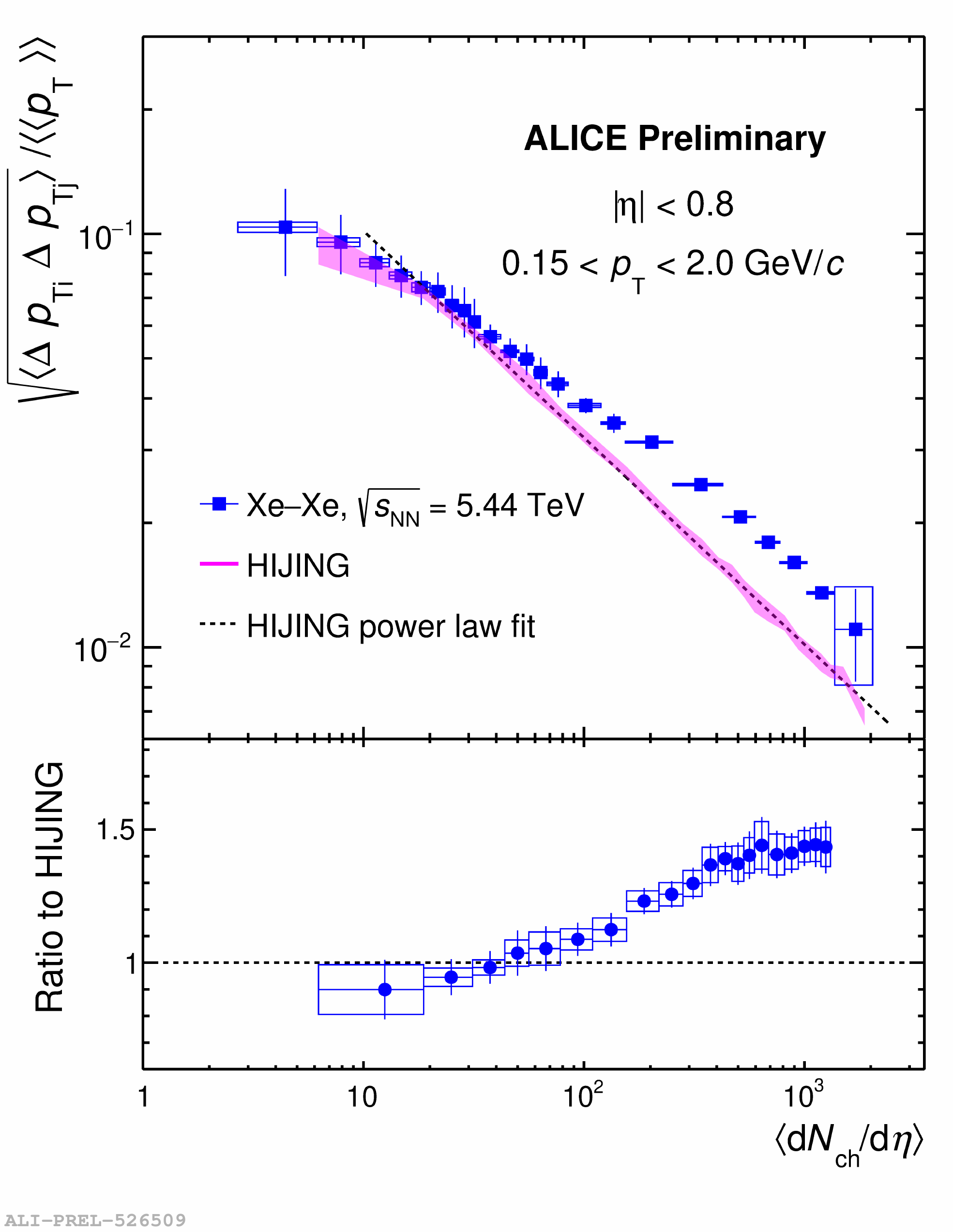}
\caption{\label{fig2} Event-by-event fluctuations of $\langle p_{\rm{T}}\rangle$ in Xe--Xe collisions at $\sqrt{s_{\rm{NN}}}$ = 5.44 TeV and comparisons with HIJING model}
\end{center}
\end{figure}

The right panel of Fig.~\ref{fig1} shows the comparison of event-by-event fluctuations of $\langle p_{\rm{T}}\rangle$ in Pb--Pb collisions at $\sqrt{s_{\rm{NN}}}$ = 5.02 TeV and Xe--Xe collisions at $\sqrt{s_{\rm{NN}}}$ = 5.44 TeV. The two results are found to be consistent with each other. Figure~\ref{fig2} shows the comparison of event-by-event $\langle p_{\rm{T}}\rangle$ fluctuations in Xe--Xe collisions at $\sqrt{s_{\rm{NN}}}$ = 5.44 TeV with HIJING model. The experimental data shows a clear deviation from the HIJING model as well as from the power law fit, indicating a deviation from a simple superposition scenario. This indicates that the final state particle production in heavy-ion collisions at the LHC can not be described by a mere superposition of independent particle-emitting sources.

\section{Conclusions}
Event-by-event $\langle p_{\rm{T}}\rangle$ fluctuations of  charged particles produced in Pb--Pb and Xe--Xe collisions at $\sqrt{s_{\rm{NN}}}$ = 5.02 TeV and $\sqrt{s_{\rm{NN}}}$ = 5.44 TeV are studied as a function of the charged-particle multiplicity. Significant dynamical fluctuations are observed in both collision systems which indicate correlated particle emission. The central Pb--Pb collisions show a significant reduction of the fluctuation in comparison to peripheral collisions. The results are compared with the HIJING model as well as with previous measurements in Pb--Pb collisions at $\sqrt{s_{\rm{NN}}}$ = 2.76 TeV. A clear deviation from a simple superposition scenario, where the final state particles are produced from a superposition of independent particle emitting sources, is observed.

\section*{References}

\end{document}